\documentclass{sig-alternate-10pt}

\usepackage{cite}
\usepackage{graphicx}
\usepackage{epstopdf}
\usepackage{epsfig}
\usepackage{amssymb}
\usepackage{amsmath}
\usepackage{multicol}
\usepackage{epsf,psfrag,pstricks}
\usepackage{rotating}
\usepackage{subfigure}

\begin{document}

\title{Characterizing User Mobility in Second Life}
\numberofauthors{2}
\author{
\alignauthor
Chi-Anh La\\
       \affaddr{Institut Eurecom}\\
       \affaddr{2229, route des Cretes}\\
       \affaddr{06560 Sophia Antipolis, France}\\
       \email{La@eurecom.fr}
\alignauthor
Pietro Michiardi\\
       \affaddr{Institut Eurecom}\\
       \affaddr{2229, route des Cretes}\\
       \affaddr{06560 Sophia Antipolis, France}\\
       \email{Pietro.Michiardi@eurecom.fr}
}

\maketitle

\begin{abstract}
In this work we present a measurement study of user mobility in Second Life. We first discuss different techniques to collect user traces and then focus on results obtained using a \emph{crawler} that we built. Tempted by the question whether our methodology could provide similar results to those obtained in real-world experiments, we study the statistical distribution of user contacts and show that from a qualitative point of view user mobility in Second Life presents similar traits to those of real humans. We further push our analysis to line of sight networks that emerge from user interaction and show that they are highly clustered. Lastly, we focus on the spatial properties of user movements and observe that users in Second Life revolve around several point of interests traveling in general short distances. Besides our findings, the traces collected in this work can be very useful for trace-driven simulations of communication schemes in delay tolerant networks and their performance evaluation.
\end{abstract}

\category{C2.1}{Computer-communication Networks}{Network Architecture and Design}



\section{Introduction}

This work is motivated by prior studies on human mobility performed in real life. For example, \cite{Karagiannis2007, Chaintreau2007, Chaintreau2007a} conduct several experiments mainly in confined areas and study analytical models of human mobility with the goal of assessing the performance of message forwarding in Delay Tolerant Networks (DTNs). Each user taking part in such experiments is equipped with a wireless device (for example a sensor device, a mobile phone, ...) running a custom software that records \emph{temporal information} about their contacts. Individual measurements are collected, combined and parsed to obtain the temporal distribution of contact times.

In this paper we present a novel methodology to capture spatio-temporal dynamics of user mobility that overcomes most of the limitations of previous attempts: it is cheap, it requires no logistic organization, it is not bound to a specific wireless technology and can potentially scale up to a very large number of participants. Our measurement approach exploits the tremendous raise in popularity of Networked Virtual Environments (NVEs), wherein thousands of users connect daily to interact, play, do business and follow university courses just to name a few potential applications. Here we focus on the SecondLife (SL) ``metaverse'' \cite{secondlife} which has recently gained momentum in the on-line community. 

Our primary goal is to perform a temporal, spatial and topological analysis of user interaction in SL. 
Prior works that attempted the difficult task of measuring and collecting traces of human mobility and contact opportunities are restricted by logistic constraints (number of participants to the experiments, duration of the experiments, failures of hardware devices, wireless technology used). In general, position information of mobile users is not available, thus a spatial analysis is difficult to achieve \cite{Chaintreau2007}. Some experiments with GPS-enabled devices have been done in the past \cite{Krumm2005, Rhee2008}, but these experiments are limited to outdoor environments.

In this paper we discuss two monitoring architectures that we tested and focus on the most robust technique, which is based on a custom software module (termed a \emph{crawler}). Our crawler connects to SL and extracts position information of all users concurrently connected to a sub-space of the metaverse: all results presented in this paper have been obtained with this architecture.

One striking evidence of our results is that they qualitatively fit to real life data, raising the legitimate question whether measurements taken in a virtual environment present similar traits to those taken in a real setting. Our methodology allows performing large experiments at a very low cost and generate data that can be used for trace-driven simulations of a large variety of applications: the study of epidemics and information diffusion in wireless networks are just some prominent examples.


\section{Monitoring architectures}\label{sec:architectures}
Mining data in a NVE can be approached under different angles. The first architecture we discuss exploits SL and its features to create objects capable of sensing user activities in the metaverse. However, there are several limitations intrinsic to this approach that hinder our ultimate goal, which is to collect a large data set of user mobility patterns. These limitations mostly come from inner design choices made by the developers of SL to protect from external attackers aiming at disrupting the system operation.
The limitations incurred by the first approach can be circumvented by building a \emph{crawler} that connects to SL as a normal user.

The task of monitoring user activity in the \textbf{whole} SL metaverse is very complex: in this work we focus on measurements made on a selected subspace of SL, that is called a land (or island). In the following we use the terminology \emph{target land} to indicate the land we wish to monitor. Lands in SL can be private, public or conceived as sandboxes and different restrictions apply: for example private lands forbid the creation and the deployment of objects without prior authorization.

We now detail the monitoring architectures we investigated in our work.

\vspace{2pt}
\noindent
\textbf{A sensor network architecture\footnote{This approach has been used also in \cite{varvello2007}.}: }Our first approach has been inspired by current research in the area of wireless sensor networks: it resembles to what one would do in the real world to measure physical data (temperature, movements, etc ...) by deploying sensor devices in the area to be monitored. We built \emph{virtual} sensors using the standard object creation tool accessible from a SL client software.  Our sensors collect data and communicate with an external web server that stores the location information of users connected to the target land. The functionality of a sensor is defined using a proprietary scripting language \cite{lsl}.

A key limitation imposed by the infrastructure of SL is that sensors cannot be arbitrarily deployed on any land. While it is impossible to deploy objects on private lands without authorization, objects on public lands expire after a predicted lifetime, which is land dependent. To deal with the restricted object lifetime, our system replicates all sensors in the same position at regular time intervals.

When a sensor is deployed on the target land, it detects users (a maximum of 16 users can be detected at the same time) that are within the sensing range (96 meters) with a tunable periodicity and stores this information in its local cache (16KB is the maximum storage space). Due to its limited memory, a sensor initiates a connection with our web server and flushes its memory using the HTTP protocol as soon as the maximal capacity has been reached. The technical specification of a sensor imposes several challenges that hinder the task of covering an entire land.
Moreover, the number of HTTP messages that can be exchanged between sensors and the web server is restricted by the SL infrastructure: this limits the quantity of data that can be retrieved from our sensors, hence a tradeoff exists between the granularity of the sensed data and the duration of a monitoring experiment. 

\vspace{2pt}
\noindent
\textbf{Monitoring using an external crawler: }An alternative approach is to build a custom SL client software (termed a \emph{crawler}) using \texttt{libsecondlife} \cite{lib}. The crawler is able to monitor the position of \textbf{every} user using a specific feature of \texttt{libsecondlife} that enables the creation of simple maps of the target land. Measurement data is stored in a database that can be queried through an interactive web application\footnote{Access to the application can be requested via mail to the authors.}. The crawler connects to the SL metaverse as a normal user, thus it is not confined by limitations imposed by private lands: any accessible land can be monitored in its totality; the maximum number of users that can be tracked is bounded only by the SL architecture (as of today, roughly concurrent 100 users per land); communication between the crawler and the database is not limited by SL.

During our experiments, we noted that introducing measurement probes in a NVE can cause unexpected effects that perturb the normal behavior of users and hence the measured user mobility patterns. Since our crawler is nothing but a stripped-down version of the legacy SL client and requires a valid login/password to connect to the metaverse, it is perceived in the SL space as an avatar, and as such may attract the attention of other users that try to interact with it: our initial experiments showed a steady convergence of user movements towards our crawler. To mitigate this perturbing effect we designed a crawler that mimics the behavior of a normal user: our crawler randomly moves over the target land and broadcasts chat messages chosen from a small set of pre-definied phrases.

\section{Measurement methodology}

Using the physical coordinates of users connected to a target land, we create snapshots of \emph{line of sight communication networks}:  given an arbitrary communication range $r$, a communication link exists two users $v_{i}, v_{j}$ if their distance is less than $r$. In the following we use a temporal sequence of networks extracted from the traces we collected using our \emph{crawler} and analyze contact opportunities between users, their spatial distribution and graph-theoretic properties of their communication network.

A precondition for being able to gather useful data is to select an appropriate target land and measurement parameters.
Choosing an appropriate target land in the SL metaverse is not an easy task because a large number of lands host very few users and lands with a large population are usually built to distribute virtual money: all a user has to do is to sit and wait for a long enough time to earn money (for free). In this work, we manually selected and analyzed the following lands: \emph{Apfel Land}, a german-speaking arena for newbies; \emph{Dance Island}, a virtual discotheque; \emph{Isle of View}, a land in which an event (St. Valentines) was organized. These lands have been chosen as they are representative of out-door (Apfel Land) and in-door (Dance Island) environments; the third land represents an example of SL events which supposedly attract many users. 
In this paper we present results for 24 hours traces: while the analysis of longer traces yields analogous results to those presented here, long experiments are sometimes affected by instabilities of \texttt{libsecondlife} under a Linux environment and we decided to focus on a set of shorter but stable measurements. A summary of the traces we analyzed can be defined based on the total number of unique users and the average number of concurrently logged in users: Isle of View had 2656 unique visitors with an average of 65 concurrent users, Dance Island had 3347 unique users and 34 concurrent users in average and Apfel Land had 1568 users and 13 concurrent users in average.

We launched the crawler on the selected target lands and set the time granularity (intervals at which we take a snapshot of the users' position) to $\tau = 10$ sec. 
We selected a communication range $r$ to simulate users equipped a bluetooth and a WiFi (802.11a at 54 Mbps) device, respectively $r_{b}=10$ meters and $r_{w}=80$ meters. In this work we assume an \emph{ideal wireless channel}: line of sight networks extracted from our traces neglect the presence of obstacles such as buildings and trees. 

User location in SL is expressed by her coordinates $\{x,y,z\}$ which are relative to the target land whose size is by default $256 \cdot 256$ meters. However there is one exception: when a user sits on an object (e.g. a bench) her coordinates are $\{x=0,y=0,z=0\}$. In the target lands we selected users did not sit.

\subsection{Temporal analysis}\label{sec:temporal}
The metrics we use to analyze mobility patterns are inspired by the work of Chaintreau \textit{et. al.} \cite{Chaintreau2007a} and allow the analysis of the statistical distribution of contact opportunities between users:

\begin{itemize}
\item \emph{Contact time $(CT)$: } is defined as the time interval in which two users $(v_{i},v_{j})$ are in direct communication range, given $r$;
\item \emph{Inter-contact time ($ICT$): } is defined as the time interval which elapses between two contact periods of a pair of users. Let $$[t_{(v_{i},v_{j})s}^1, t_{(v_{i},v_{j})e}^1],[t_{(v_{i},v_{j}j)s}^2, t_{(v_{i},v_{j})e}^2],...[t_{(v_{i},v_{j})s}^n, t_{(v_{i},v_{j})e}^n]$$ be the successive time intervals at which a contact between user $v_{i}$ and $v_{j}$ occurs; then, the inter-contact time between the $k-th$ and the $(k + 1)-th$ contact intervals is: $$ICT_{(v_{i},v_{j})}^k = t_{(v_{i},v_{j})s}^{k+1} - t_{(v_{i},v_{j})e}^{k} $$ 
\item \emph{First contact time ($FT$): } is defined as the waiting time for a user $v_{i}$ to contact her first neighbor (ever).
\end{itemize}

\subsection{Spatial analysis}\label{sec:spatial}
We present here the metrics we used to perform the spatial analysis of our traces:

\begin{itemize}
\item \emph{Node degree: } is defined as the number of neighbors of a user when the communication range is fixed to $r$;
\item \emph{Network diameter: } is computed as the longest shortest path of the largest connected component of the communication network formed by the users. We used the largest component since, for a given $r$, the network might be disconnected;
\item \emph{Clustering coefficient: } is defined as in \cite{Watts1998}: we compute it for every user and take the mean value to be representative of the whole communication network;
\item \emph{Travel length: } for every user $v_{i}$ we compute the distance covered from its login to its logout coordinates in SL;
\item \emph{Effective Travel time: } for every user $v_{i}$ we compute the total time spent while moving; hence, this metric does not include \emph{pause} times; 
\item \emph{Travel time: } for every user $v_{i}$ we compute the total connection time to the SL land we monitor with the crawler;
\item \emph{Zone occupation: } we divided lands in several square sub-cells of size $LxL$ and computed the number of users in every sub-cell, when $L=20$ meters.
\end{itemize}

\section{Results}
\begin{figure*}[ht!]
 \begin{tabular}{ccc}
   \begin{minipage}[t]{0.3\textwidth}
     \begin{center}

\subfigure[]{\includegraphics[scale=0.27]{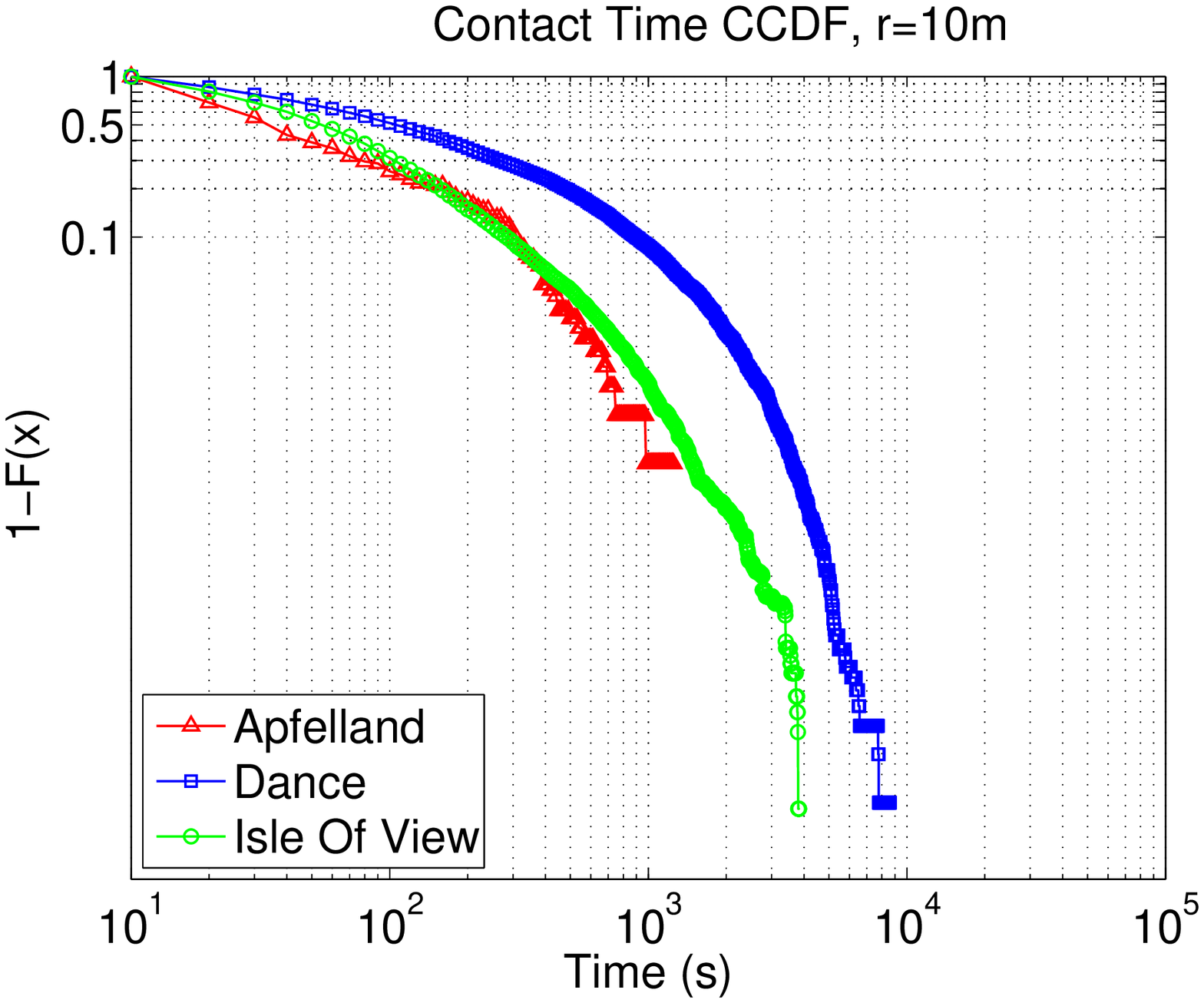}}
\label{fig:CT}

	\end{center}
   \end{minipage}
   &
   \begin{minipage}[t]{0.3\textwidth}
     \begin{center}
\subfigure[]{\includegraphics[scale=0.27]{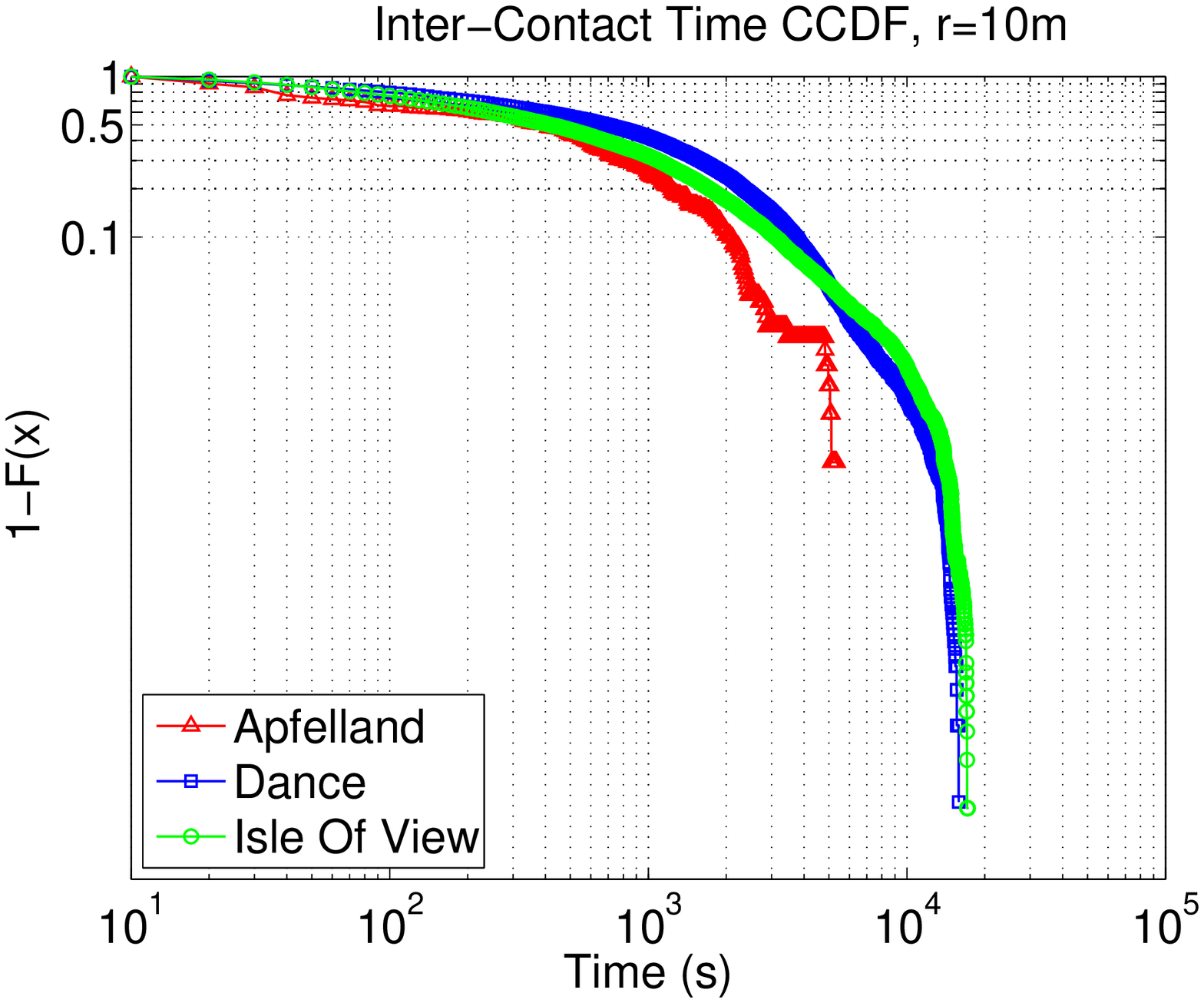}}
\label{fig:FCT}
\end{center}
   \end{minipage}
   &
   \begin{minipage}[t]{0.3\textwidth}
     \begin{center}
\subfigure[]{\includegraphics[scale=0.27]{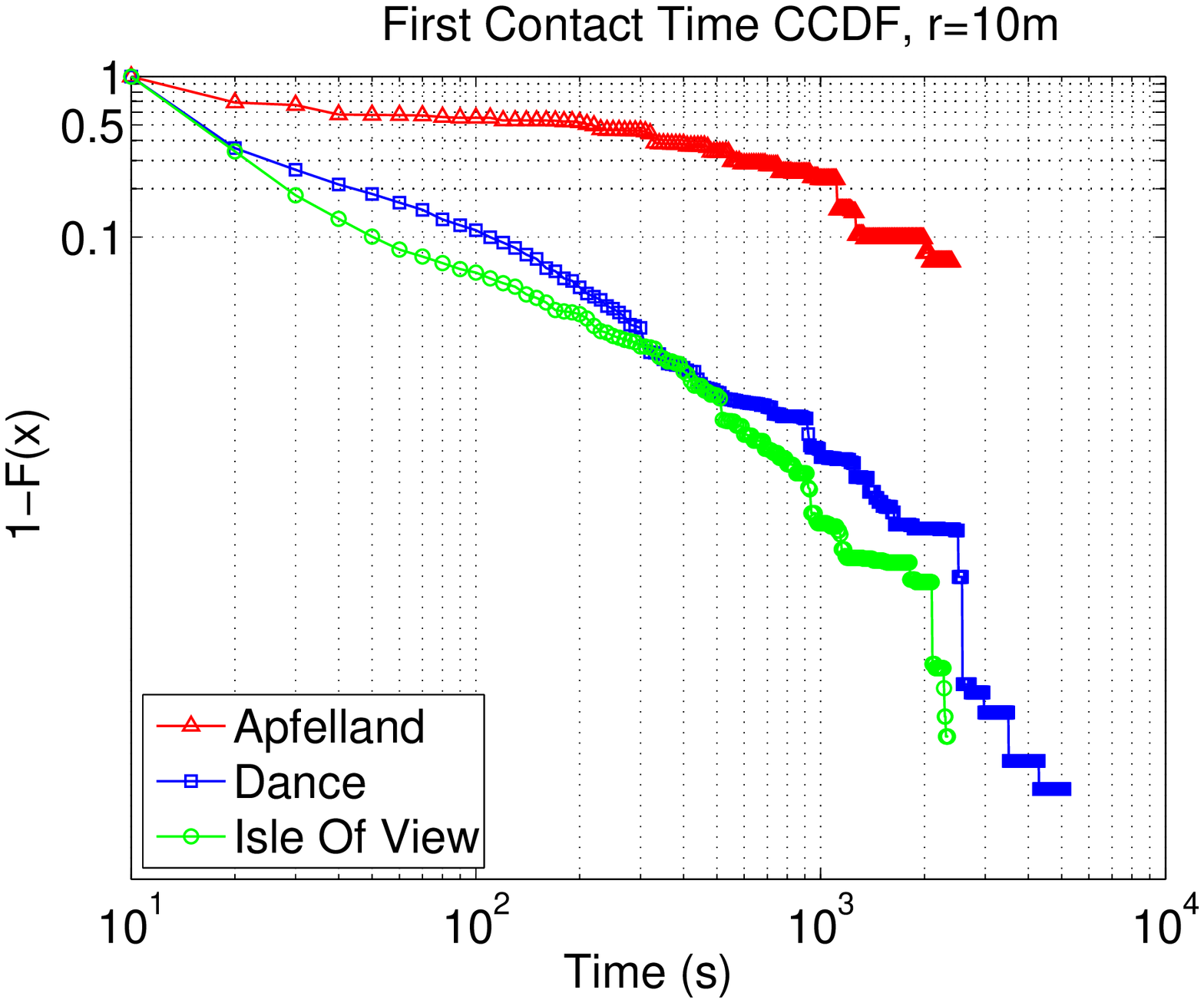}}
\label{fig:ICT}
\end{center}
   \end{minipage}

\\

\begin{minipage}[t]{0.3\textwidth}
     \begin{center}

\subfigure[]{\includegraphics[scale=0.27]{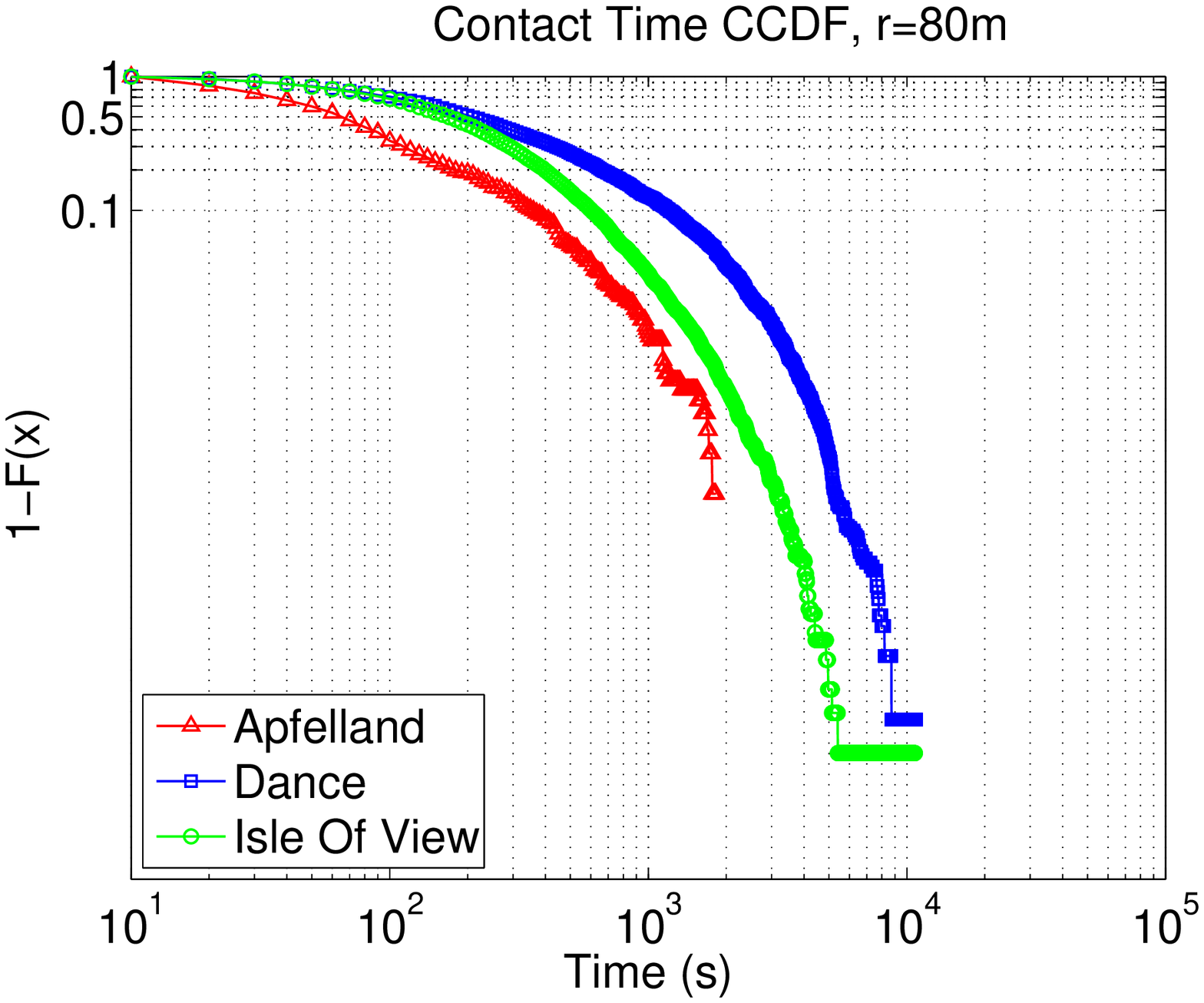}}
\label{fig:CT}

	\end{center}
   \end{minipage}
   &
   \begin{minipage}[t]{0.3\textwidth}
     \begin{center}
\subfigure[]{\includegraphics[scale=0.27]{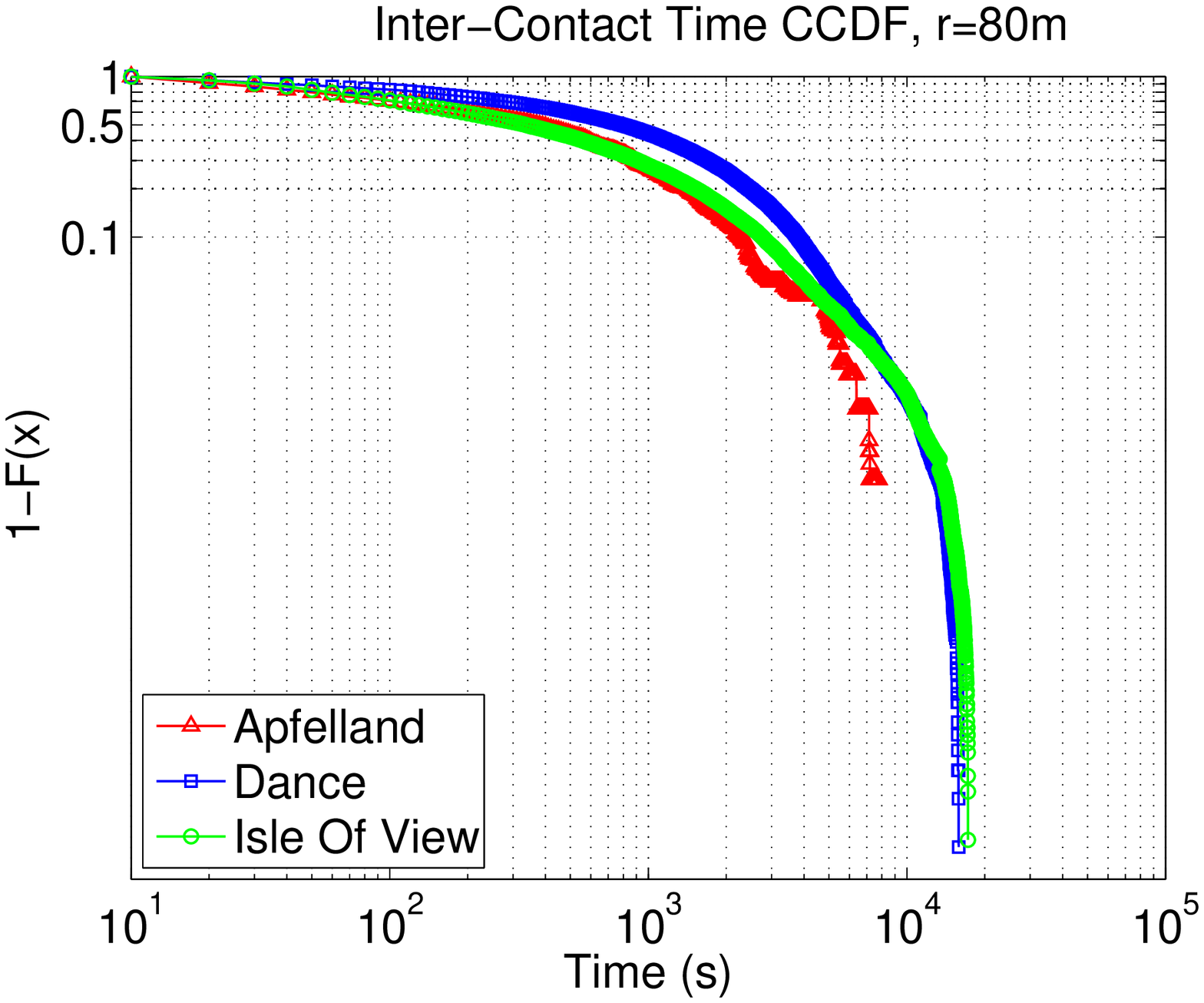}}
\label{fig:FCT}
\end{center}
   \end{minipage}
   &
   \begin{minipage}[t]{0.3\textwidth}
     \begin{center}
\subfigure[]{\includegraphics[scale=0.27]{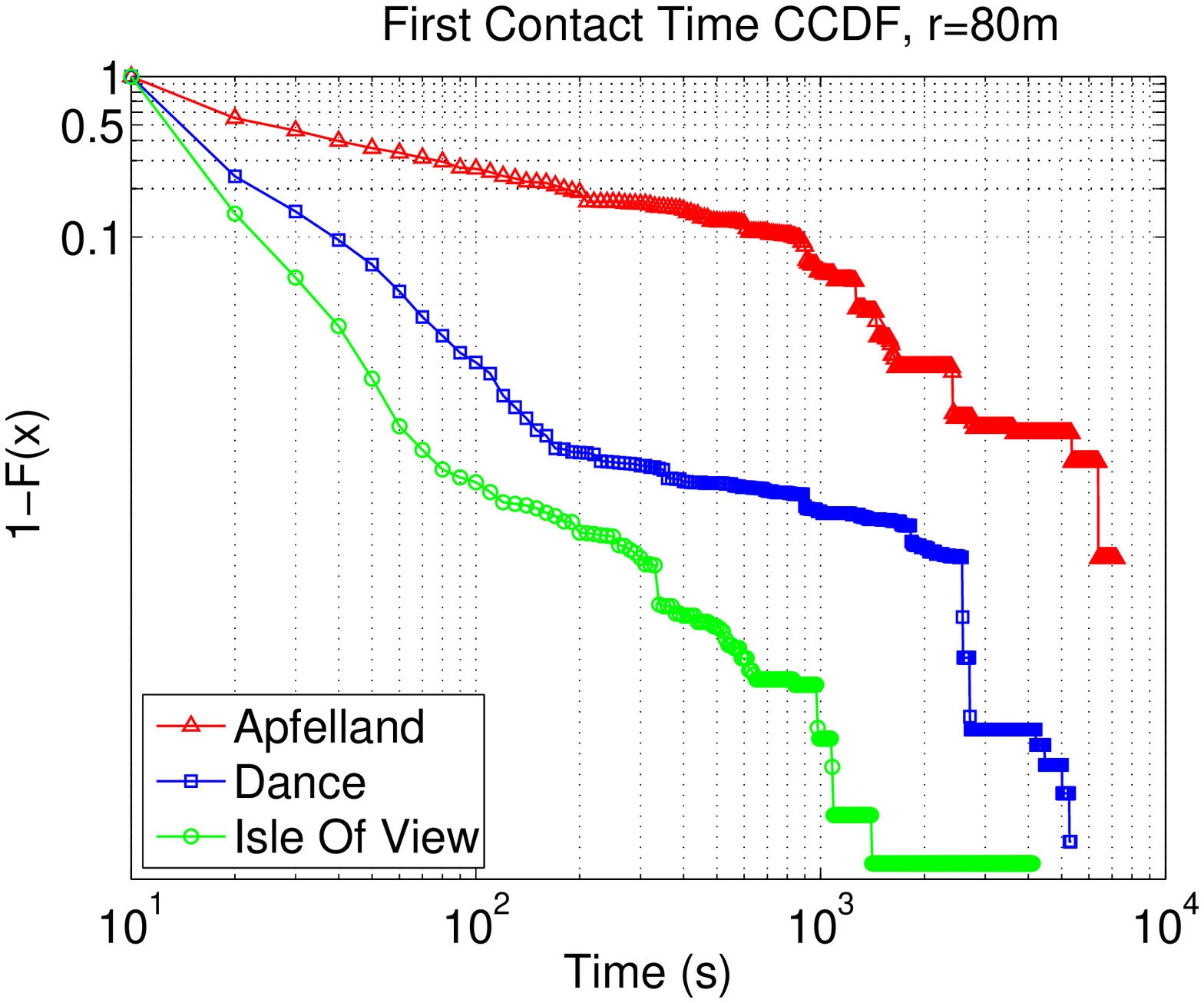}}
\label{fig:ICT}
\end{center}
   \end{minipage}
     \end{tabular}
\caption{\footnotesize{\emph{Temporal Analysis:} Complementary CDF of contact opportunity metrics for three target lands.}}
\label{fig:temporal}
\end{figure*}

We now discuss the results of our measurements for the three selected target lands and study the influence of the communication range ($r_{b}$ or $r_{w}$).

\vspace{2pt}
\noindent
\textbf{Temporal Analysis: }Fig.~\ref{fig:temporal} illustrates the distribution of the temporal metrics we used in this work for $r_{b}=10$ meters and $r_{w}=80$ meters. A glance at the complementary CDF (CCDF) of the contact time $CT$ indicates that the \emph{median} contact time is roughly 30, 60 and 100 seconds respectively for Apfel Land, Isle of View and Dance Island when $r=r_{b}$, and about 70, 200 and 300 seconds for the same set of islands when $r=r_{w}$. Besides the intuitive result which indicates larger transfer opportunities for larger $r$, we observe that the distribution of $CT$ has two phases: a first power-law phase and an exponential cut-off phase that limits the $CT$ to a few hundreds seconds.

Similar observations can be done for the CCDF of the inter contact time $ICT$: for the three target lands we analyzed, the distribution follows a first power-law phase, followed by an exponential cut-off phase. The median $ICT$ is around 400 seconds for the two open-space lands and between 700 and 800 seconds for the Dance Island. Analyzing the same trace of user movement yields surprisingly similar results with different communication ranges. We believe this result is due to the fact that users are concentrated around point of interests (as discussed below), but it would be interesting to compare such findings with real-world experiments.

Although the distribution of contact opportunities appears to be similar for the two open-space lands, the CCDF of the first contact time $FT$ illustrates some differences between these lands: in Apfel Land users have to wait for a long time before meeting their first neighbor. The median $FT$ is around 300 seconds for Apfel Land, while it is less than 20 seconds for the other two lands when $r=r_{b}$. The $FT$ improves a lot when increasing $r$: the median is around 30 seconds for Apfel Land and less than 5 seconds for the other lands.

These results are quite surprising: from a \emph{qualitative} point of view, we obtained a statistical distribution of contact opportunities that mimics what has been obtained for experiments in the real world \cite{Krumm2005, Chaintreau2007, Rhee2008}. Obviously, human activity roughly spans the 12 hours interval, while even the most assiduous user which we were able to track spent less than 4 consecutive hours on SL, hence a quantitative comparison is not immediate. 

\begin{figure*}[ht!]
 \begin{tabular}{ccc}
   \begin{minipage}[t]{0.3\textwidth}
     \begin{center}

\subfigure[]{\includegraphics[scale=0.3]{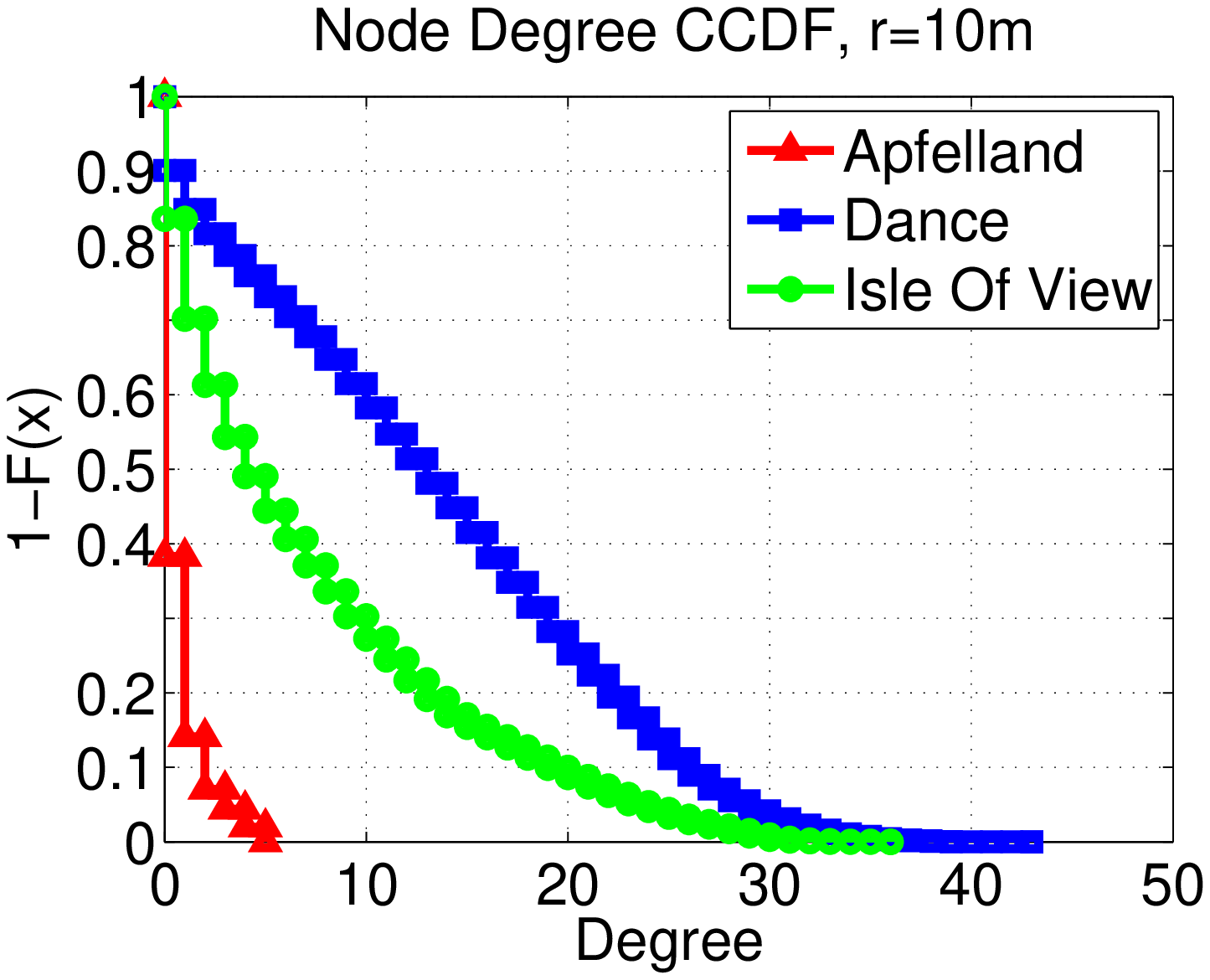}}
\label{fig:CT}

	\end{center}
   \end{minipage}
   &
   \begin{minipage}[t]{0.3\textwidth}
     \begin{center}
\subfigure[]{\includegraphics[scale=0.3]{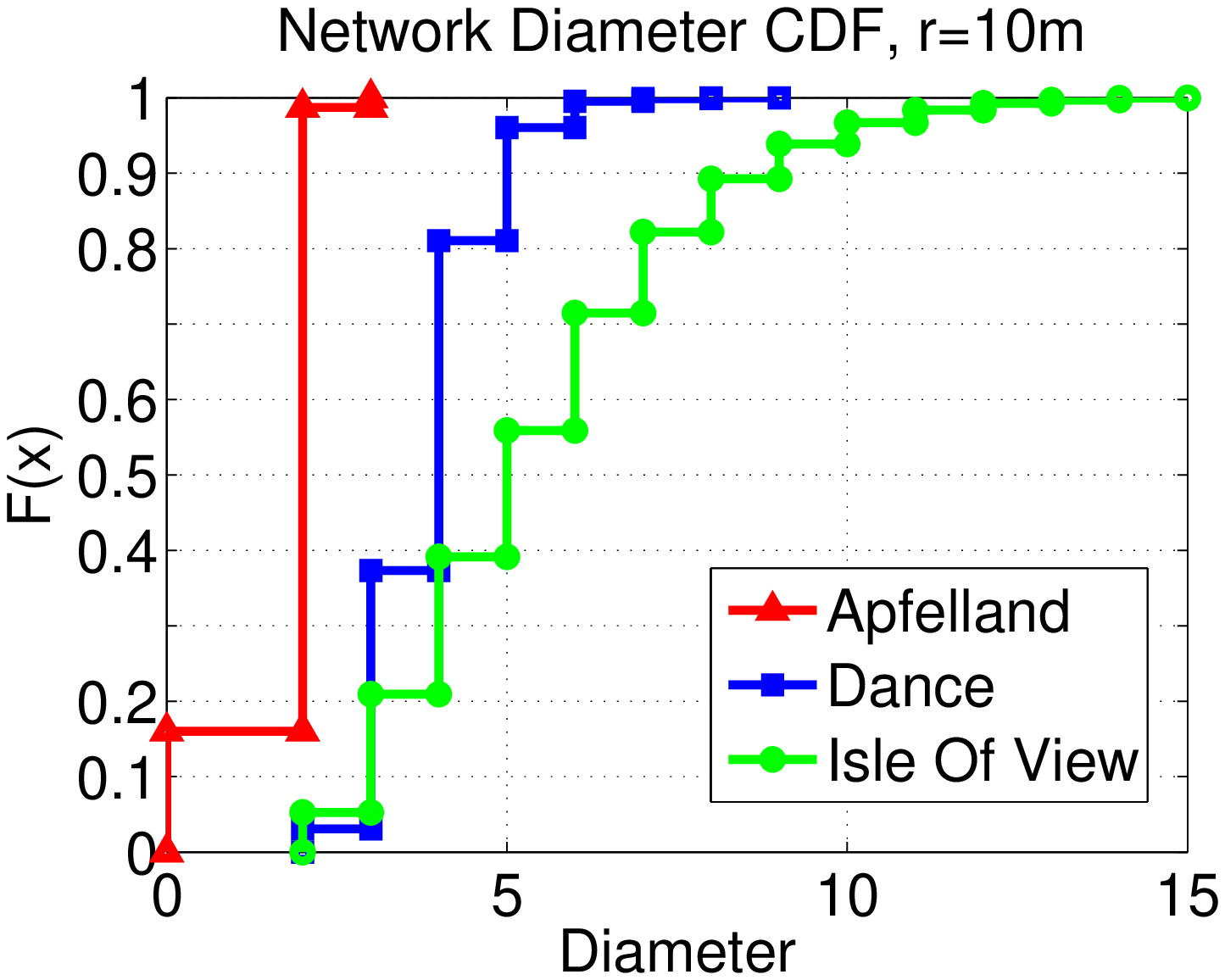}}
\label{fig:FCT}
\end{center}
   \end{minipage}
   &
   \begin{minipage}[t]{0.3\textwidth}
     \begin{center}
\subfigure[]{\includegraphics[scale=0.3]{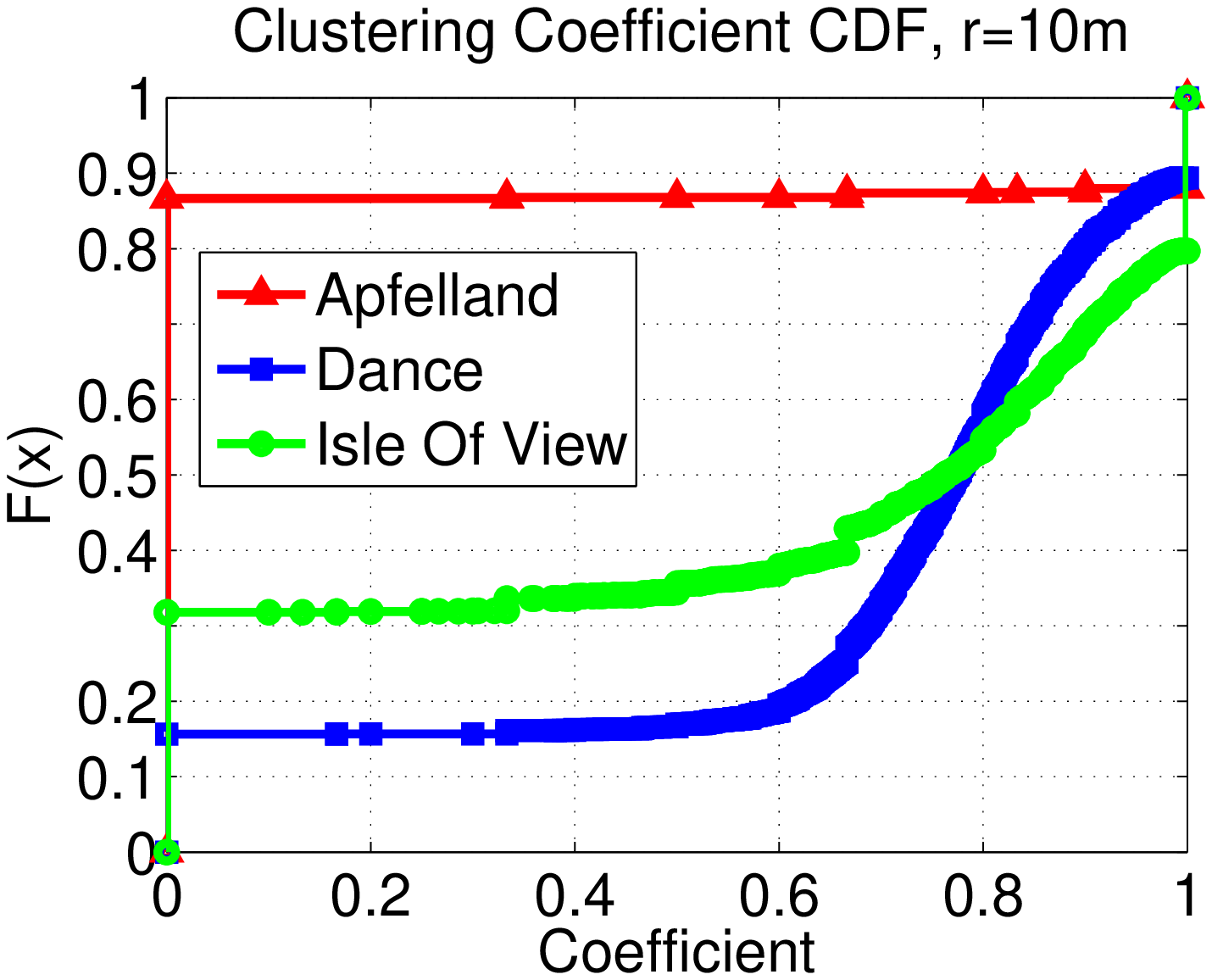}}
\label{fig:ICT}
\end{center}
   \end{minipage}

\\

\begin{minipage}[t]{0.3\textwidth}
     \begin{center}

\subfigure[]{\includegraphics[scale=0.3]{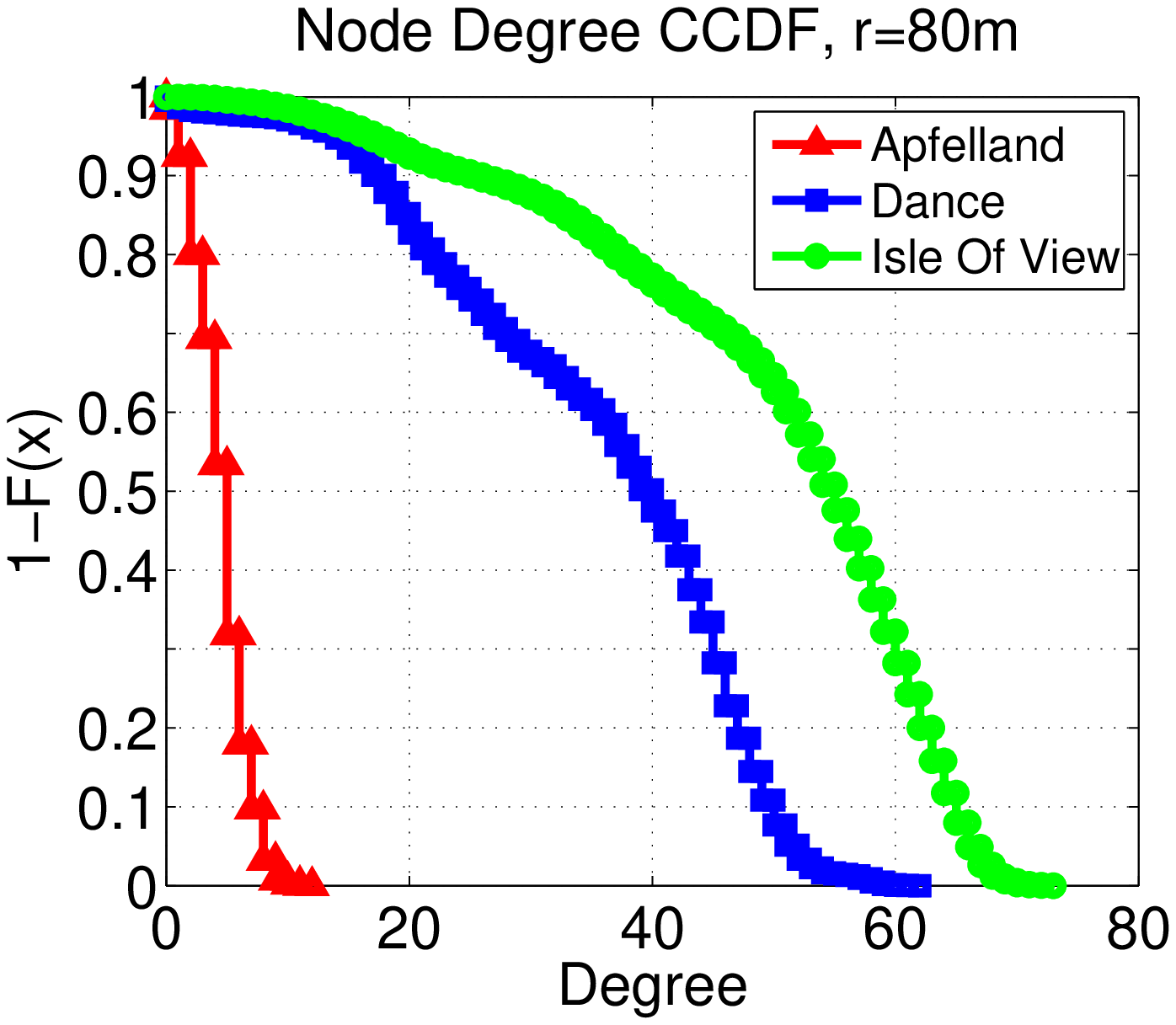}}
\label{fig:CT}

	\end{center}
   \end{minipage}
   &
   \begin{minipage}[t]{0.3\textwidth}
     \begin{center}
\subfigure[]{\includegraphics[scale=0.3]{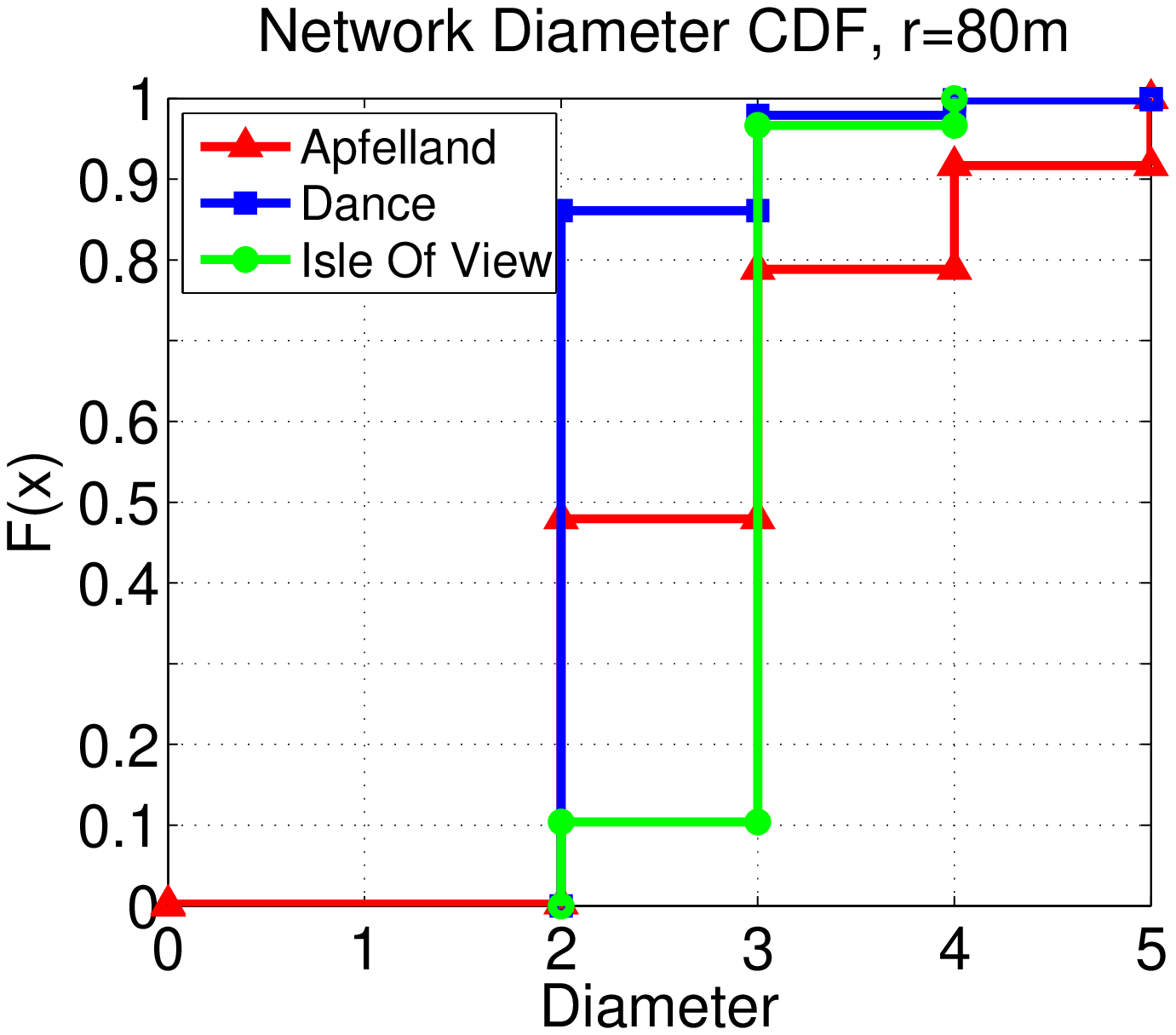}}
\label{fig:FCT}
\end{center}
   \end{minipage}
   &
   \begin{minipage}[t]{0.3\textwidth}
     \begin{center}
\subfigure[]{\includegraphics[scale=0.3]{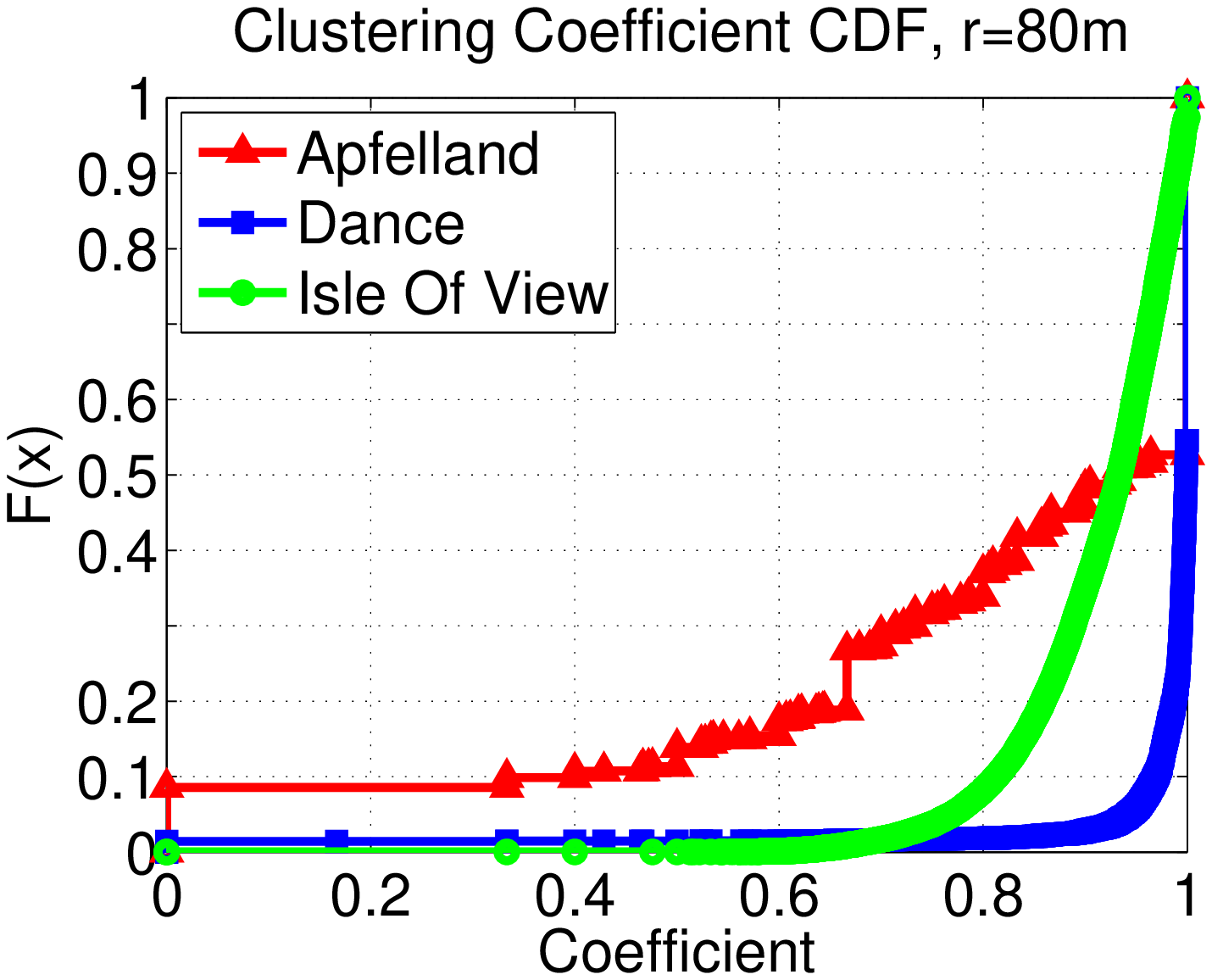}}
\label{fig:ICT}
\end{center}
   \end{minipage}
     \end{tabular}
\caption{\footnotesize{\emph{Line of sight networks:} graph theoretic properties for three selected target lands.}}
\label{fig:graph}
\end{figure*}

\begin{figure}[htb!]
\begin{center}
\includegraphics[scale=0.27]{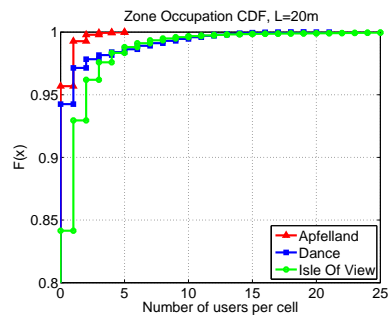}
\caption{\footnotesize{\emph{Spatial distribution} of users.}}
\label{fig:spatial}
\end{center}
\end{figure}

\vspace{2pt}
\noindent
\textbf{Line of sight networks: } We now delve into a detailed analysis of the communication networks that emerge from user interaction when we assume them to be equipped with a wireless communication device covering a range $r \in \{ r_{b},r_{w} \}$. Fig.~\ref{fig:graph} illustrates the aggregated (over the whole measurement period) CCDF of the node degree, the aggregated CDF of the network diameter and clustering coefficient.

The node degree CCDF illustrates a diverse user behavior in each target land: for Apfel Land we observe that 60\% of users have no neighbors, for the Dance Island only 10\% of users have no neighbors while in the Isle of View, all users have at least one neighbor when $r=r_{b}$. When the communication range is set to $r=r_{w}$ all users have at least one neighbor in all lands. The maximum degree and the whole distribution varies a lot between target lands: the main reason lies in the physical distribution of users on a land. In Apfel Land users are relatively sparse while in the Dance Island, for example, most of the users spend most of the time in a tiny portion of the land: this observation is corroborated\footnote{There is an intuitive reason for this phenomenon: in a discotheque users spend most of their time on the dance floor or by the bar, while in an open space users are generally located more sparsely.} by our study on the spatial distribution of users as shown in Fig.~\ref{fig:spatial}. Although the general trend for all target lands we inspected is that a large fraction of the land has no users, some lands (e.g. Dance Island) are characterized by hot-spots with several tens of users.

The CDF of the network diameter illustrates the impact of different transmission ranges: it is clear that the diameter shrinks for $r=r_{w}$. We note, however, that for Apfel Land there is an apparent contradiction: for $r=r_{b}$ the maximum diameter is smaller than for $r=r_{w}$. This phenomenon is due to the fact we compute the diameter of the largest connected component of the temporal graph formed by users: when the radio range is small (and users are scattered through the target land) we observe the emergence of relatively small connected components, whereas for larger ranges the connected component is large (eventually it includes all users), hence a larger diameter.

In Fig.~\ref{fig:spatial} we also plot the CDF of the clustering coefficient for the whole measurement period. Our results clearly point to high \emph{median} values of the clustering coefficient which indicate that the networks we observe are not random graphs\footnote{Which are usually characterized by a very small clustering coefficient \cite{Watts1998}.}: these networks are highly clustered but, due to the small number of concurrent users that can log in to a land and the results on the network diameter, we cannot claim at this time that the graphs that emerge from user interaction have small world characteristics.

\begin{figure*}[ht!]
 \begin{tabular}{ccc}
   \begin{minipage}[t]{0.3\textwidth}
     \begin{center}

\subfigure[]{\includegraphics[scale=0.27]{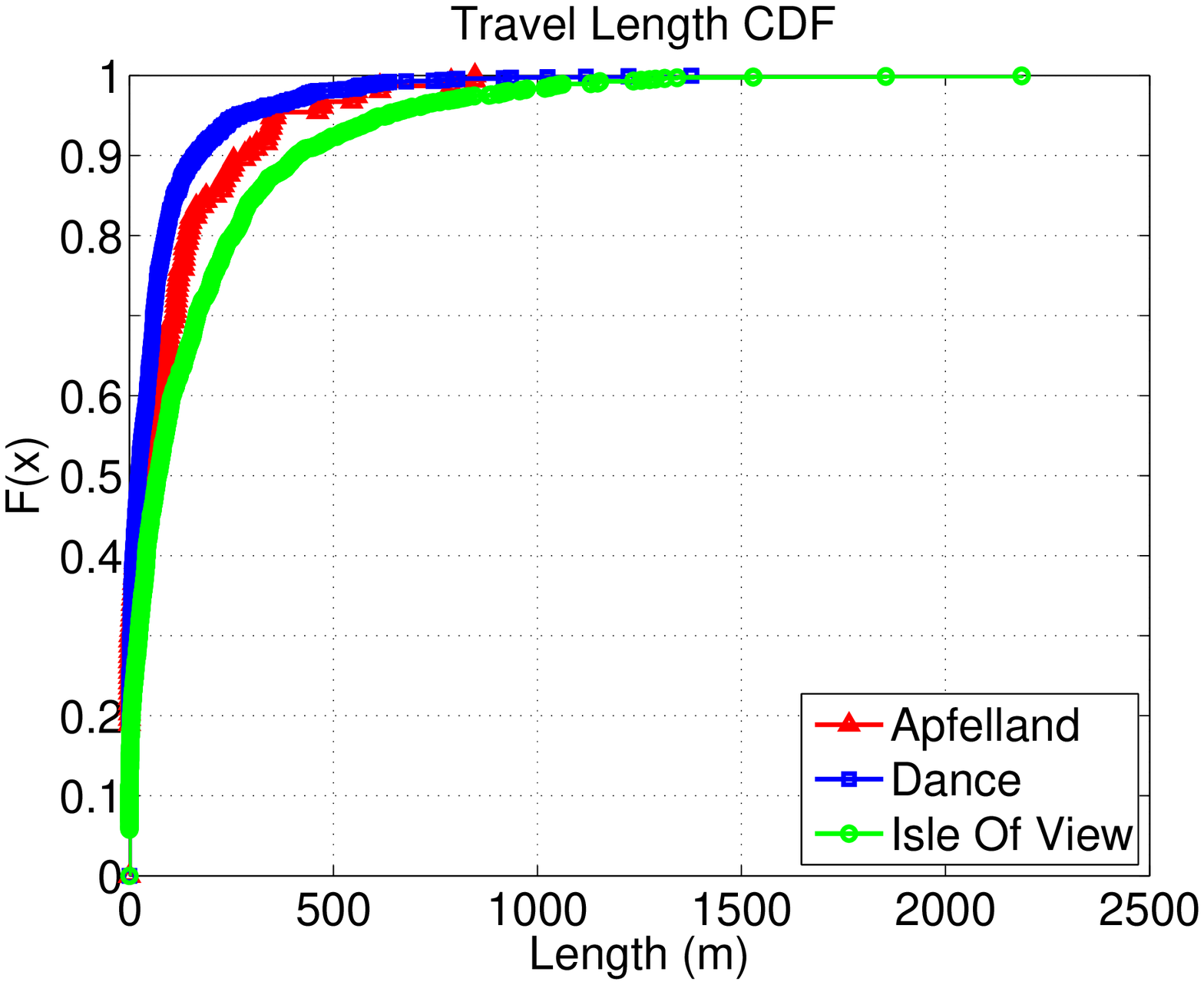}}
\label{fig:CT}

	\end{center}
   \end{minipage}
   &
   \begin{minipage}[t]{0.3\textwidth}
     \begin{center}
\subfigure[]{\includegraphics[scale=0.27]{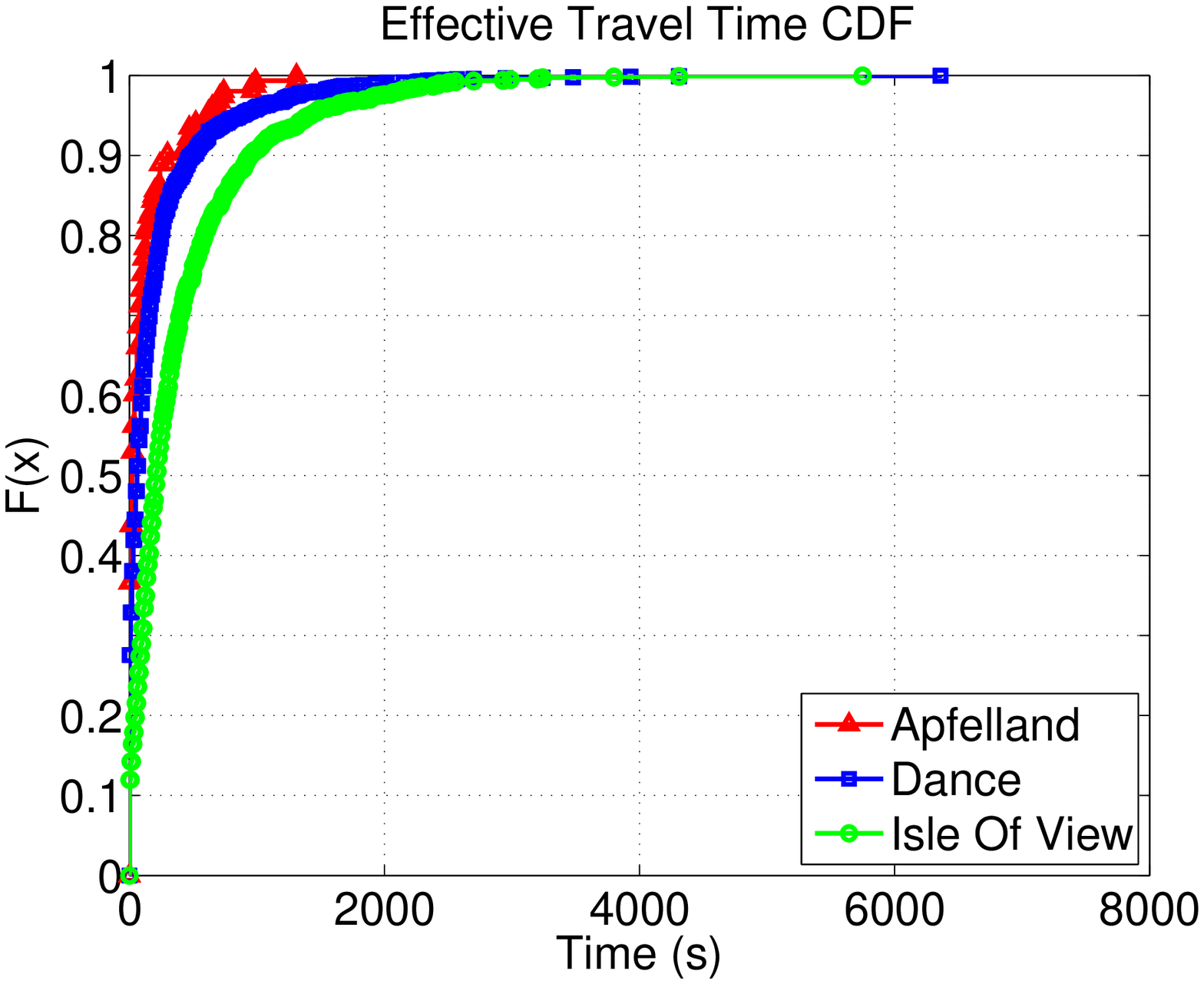}}
\label{fig:FCT}
\end{center}
   \end{minipage}
   &
   \begin{minipage}[t]{0.3\textwidth}
     \begin{center}
\subfigure[]{\includegraphics[scale=0.27]{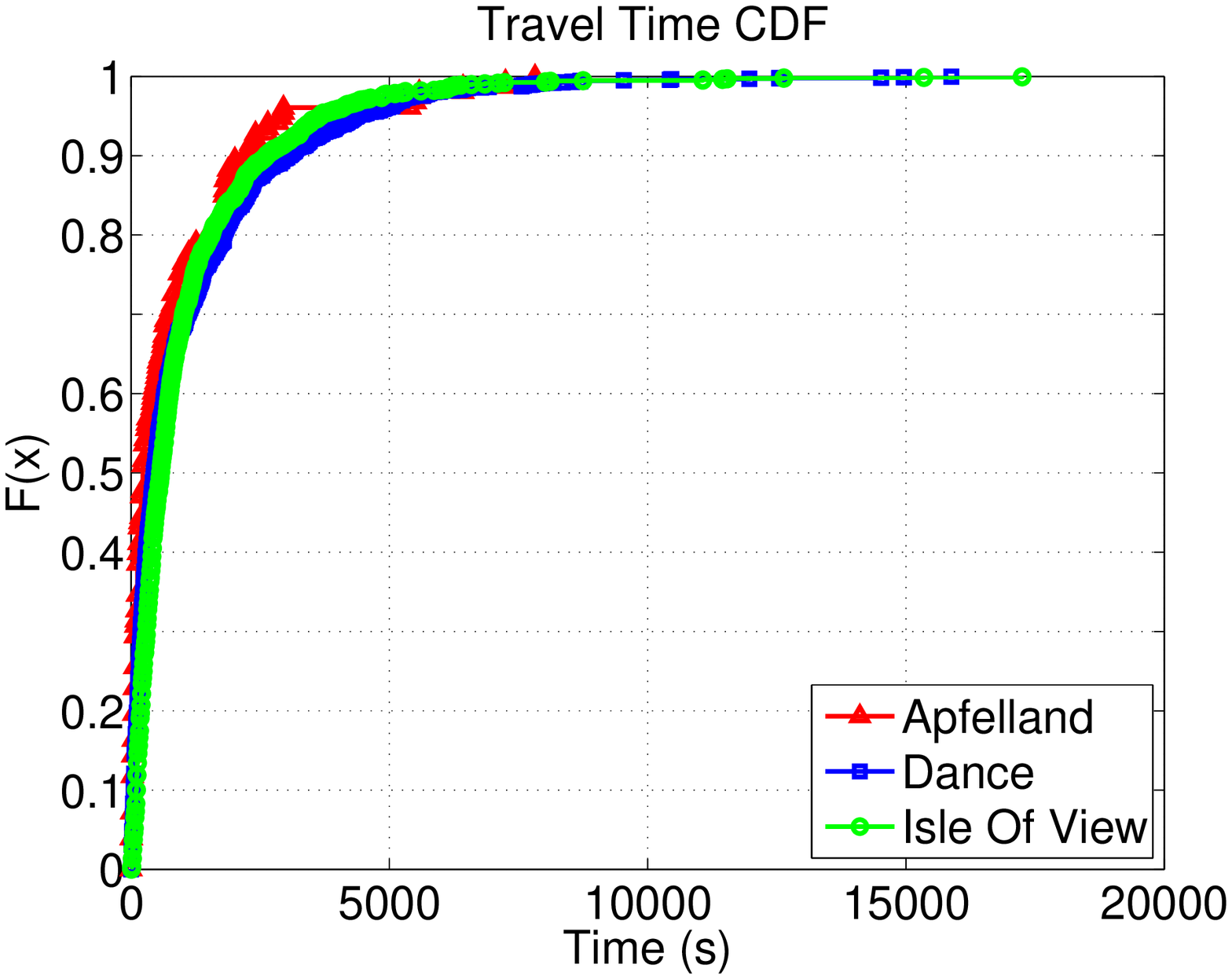}}
\label{fig:ICT}
\end{center}
   \end{minipage}
     \end{tabular}
\caption{\footnotesize{\emph{Trip analysis} for three selected target lands}}
\label{fig:trip}
\end{figure*}

\vspace{2pt}
\noindent
\textbf{Trip analysis: } using physical coordinates, we were able to study the statistical distribution of the distance travelled by users on the three target lands we analyze in this paper. Fig.~\ref{fig:trip} illustrates the aggregate CDF of the travel length, the travel time and the login time for all users. Fig.~\ref{fig:trip}(c) shows the CDF of the login time: in our measurement we observed that the longest log-in time for a user was around 4 hours while 90\% of users are logged in for less than 1 hour. 

Fig.~\ref{fig:trip}(a) provides further hints towards a better understanding of user mobility in the selected target lands. For a confined area such as Dance Island, the vast majority of users travel less than 230 meters ($90th$ percentile). This observation however applies also for open spaces: for Apfel Land, the $90th$ percentile is around 400 meters while it grows up to 500 meters for Isle of View. There is a small fraction of users who travel a very long distance: for the Isle of View, around 2\% of users travel more than 2000 meters. Fig.~\ref{fig:trip}(b) is useful to infer the distribution of the times a user takes to travel from her initial point (the first time our crawler tracked the user) to her final point (the last time the user has been seen on the target land). 

\section{Conclusion and future work}
In this paper we discussed a novel methodology to perform user profiling that exploits the raising popularity of on-line communities emerging from user interaction in Networked Virtual Environments. We studied the mobility patterns of users connected to Second Life using a crawler that extracts at regular time intervals user position on a target land. Tempted by the question whether any similarity can be found between our results and measurements performed in the real world, we first characterized the statistical distribution of contact opportunities among users. Our analysis indicated that mobility patterns in a virtual environment share common traits, from a qualitative point of view, with those in the real world. We further pushed our analysis to characterize the spatial distribution of users and their mobility behavior: users are generally concentrated around points of interest and travel small distances in the vast majority of cases. Finally we characterized the graph theoretic properties of line of sight networks emerging from user interaction and found results indicating they are highly clustered.

Our measurements are publicly available and constitute a useful material for trace-driven simulations of a large variety of applications: the study of epidemics and information diffusion in wireless networks, the performance analysis of forwarding schemes in DTNs, etc...

\emph{Is mobility of users in SL representative of real human mobility?} In our future work we will try and address this question from a qualitative point of view. In this paper we have constructed a tool that helps answering this key question, but we believe that further study in the specification of new metrics to define human mobility are required.
Another interesting area of future research would be to build the network of ``relationships'' among SL users. Based on the ``relation graph'', new questions can be addressed such as the frequency and the strength of contact between acquaintances.

\bibliographystyle{abbrv}
\bibliography{rr08212}

\begin{thebibliography}{10}

\bibitem{lib}
Libsecondlife: www.libsecondlife.org/.

\bibitem{lsl}
Lsl: http://wiki.secondlife.com/wiki/lsl\_portal.

\bibitem{secondlife}
Second life: http://www.secondlife.com.

\bibitem{Chaintreau2007a}
A.~Chaintreau, P.~Hui, J.~Crowcroft, C.~Diot, J.~Scott, and R.~Gass.
\newblock Impact of human mobility on opportunistic forwarding algorithms.
\newblock {\em IEEE Transactions on Mobile Computing}, 2007.

\bibitem{Chaintreau2007}
A.~Chaintreau, A.~Mtibaa, L.~Massoulie, and C.~Diot.
\newblock The diameter of opportunistic mobile networks.
\newblock In {\em Proc. of CoNEXT}, 2007.

\bibitem{Karagiannis2007}
T.~Karagiannis, J.-Y.~L. Boudec, and M.~Vojnovic\'.
\newblock Power law and exponential decay of inter contact times between mobile
  devices.
\newblock In {\em Proc. of MOBICOM}, 2007.

\bibitem{Krumm2005}
J.~Krumm and E.~Horvitz.
\newblock The microsoft multiperson location survey.
\newblock Technical report, Microsoft Research, MSR-TR-2005-13, 2005.

\bibitem{Rhee2008}
I.~Rhee, M.~Shin, S.~Hong, K.~Lee, and S.~Chong.
\newblock On the levy-walk nature of human mobility.
\newblock In {\em Proc. of IEEE INFOCOM}, 2008.

\bibitem{varvello2007}
M.~Varvello, E.~Biersack, and C.~Diot.
\newblock A networked virtual environment over kad.
\newblock In {\em Proc. of ACM CoNext}, 2007.

\bibitem{Watts1998}
D.~J. Watts and S.~Strogatz.
\newblock Collective dynamics of 'small-world' networks.
\newblock {\em Nature}, 1998.

\end{thebibliography}

\end{document}